\documentclass[prb,aps,twocolumn,showpacs,floats]{revtex4}
\usepackage{graphicx}

\begin{document}

\title{Statistics of local density of states in \\
the Falicov-Kimball model with local disorder}

\author{Minh-Tien Tran}
\affiliation{Asia Pacific Center for Theoretical Physics, Pohang, Republic of Korea, and \\
Institute of Physics and Electronics, Vietnamese Academy of Science and Technology,
 Hanoi, Vietnam.
}

%\maketitle
%\date{\today}
%\widetext\end{abstract}
%\vskip 0.1 truein
\pacs{71.27.+a, 71.23.An, 71.30.+h, 71.10.Fd}

\begin{abstract}
Statistics of the local density of states in the two-dimensional Falicov-Kimball model 
with local disorder is studied by employing the statistical dynamical mean-field theory.
Within the theory the local density of states and its distributions are calculated through
stochastic self-consistent equations. The most probable value of the local density of states 
is used to monitor the metal-insulator transition driven by correlation and disorder.
Nonvanishing of the most probable value of the  local density of states at the Fermi energy
indicates the existence of extended states in the two-dimensional disordered interacting system.  
It is also found that
the most probable value of the local density of states exhibits a discontinuity when
the system crosses from extended states to the Anderson localization.
A phase diagram is also presented. 
\end{abstract}

%\vskip2pc]

\maketitle

\section{Introduction}

Electron interaction and disorder strongly influence the properties of materials. In particular, the motion
of charge carrier particles can be suppressed by Coulomb interaction and disorder, and the suppression leads to a metal-insulator transition (MIT). In the pure system without disorder the MIT can occur and is purely driven by electron correlations.\cite{Mott} The transition is commonly referred to the Mott-Hubbard MIT. In the presence of disorder, the MIT can occur even without electron interactions. The state of system changes from extended phase to the Anderson localization due to coherent backscattering from randomly distributed
impurities.\cite{Anderson} The Mott-Hubbard MIT is characterized by opening a gap of the density of states 
(DOS) at
the Fermi energy, while the Anderson localization is a gapless insulator. At the Anderson localization the spectra change from continuous to dense discrete points. It is characterized by vanishing of the most
probable value of the local density of states (LDOS).\cite{Anderson,Anderson1,Anderson2} 
The most probable value of the LDOS discriminates 
between the metal and insulator phases. The importance of the distribution
of the LDOS in the proper description of the Anderson localization was stressed by 
Anderson.\cite{Anderson,Anderson1,Anderson2} The very distribution
of the LDOS determines the Anderson localization but not average quantities of the LDOS.  
From the distribution of the LDOS one can detect both the vanishing of the most probable value of LDOS as well as the gap opening of the total DOS, thus the distribution of the LDOS is a valuable tool for determining 
the MIT driven by both disorder and correlation.       

In recent years, a generalized Curie-Weiss mean-field theory, the dynamical 
mean-field theory (DMFT) was developed.\cite{GKKR} The DMFT essentially captures local temporal
fluctuations. It has been widely applied to study correlated electron systems. 
The DMFT describes the Mott-Hubbard MIT well.\cite{GKKR}
However, it works with the arithmetic average of the LDOS, and cannot 
determine the Anderson localization in disordered systems.
A statistical variant of the DMFT, which is usually
referred to the statistical DMFT, has been introduced to study systems with both disorder and 
interaction.\cite{Dobrosavljevic,Dobrosavljevic2}
It can be viewed as the DMFT formulated in 
real space with general inhomogeneous solutions.\cite{Tran} Within the statistical DMFT, the
self energy is a local function of frequency, but it also depends on the site index.
In the presence of diagonal disorder, the self energy is also diagonal random variables,
and gives additional dynamical random contributions to disorder.
It generates a set of self-consistent stochastic equations.
The statistical DMFT essentially deals with the LDOS, hence it is capable of studying the 
Anderson-Mott-Hubbard MIT. In parallel with the statistical DMFT, a typical medium theory (TMT) was  
also introduced to study
the Anderson localization.\cite{Dobrosavljevic1} The TMT is based on the DMFT too. However, instead of
the arithmetic average DOS, it works with the geometric
average DOS. The geometric average DOS is incorporated into the self-consistent stochastic DMFT equations,
which result into  self-consistent equations of the DMFT fashion. The TMT is essentially a mean-field
theory of both disorder and correlation, while in the statistical DMFT disorder is treated exactly, and
only the correlation effects are treated in a mean-field manner. 
The TMT was employed to study the Anderson-Mott-Hubbard MIT in correlated electron systems with local
disorder.\cite{Byczuk,Byczuk1}

In the presence of disorder the LDOS forms a stochastic ensemble. The stochastic ensemble of the LDOS must have
characteristics which discriminate between the metallic and insulator phases, and one has to search the
characteristics for determining the MIT. Examples for the characteristics are the most probable value
of the LDOS in the original Anderson theory of localization\cite{Anderson,Anderson1,Anderson2} or
the geometric average of the LDOS in the TMT.\cite{Dobrosavljevic1} 
However, nearby the critical point of the MIT, quantum fluctuations may induce outliers of the statistical
description of the stochastic ensemble of the LDOS. 
For a random sample an estimator is called robust if it is insensitive to outliers.\cite{Bickel}
The most probable value, arithmetic average, median are the examples of robust estimator.
The geometric average is not a robust estimator because it does not fulfil the linear property of the
robust estimator.\cite{Bickel} 
From the point of view of robust statistics the most probable value is a better estimator of a random sample 
than the geometric average. Moreover, in general, the geometric average DOS is not the most
probable value of the LDOS, hence it does not truly represent a typical DOS, although it is closer to the most
probable value of the LDOS than the arithmetic average of the LDOS. The geometric average DOS is sensitive
to small values of the LDOS at individual sites, even when these values do not represent the most probable value
of the LDOS. A decline of the geometric average 
DOS with increasing the disorder strength does not necessarily imply the approach to the Anderson localization.\cite{Song} Nevertheless, when the geometric average is embedded into the self-consistent
cycle of the DMFT, the whole method, i.e the TMT, can describe the Anderson 
localization.\cite{Dobrosavljevic1,Byczuk,Byczuk1}  

The interplay between disorder and correlation in the MIT theory is a long standing problem. In the 
clean or noninteracting limits the MIT was well studied.\cite{GKKR,Lee} However, 
the correlation effects in disordered systems still remain unclear. In particular, 
despite experiments found evidences of the metallic behavior in the
two dimensional electron systems, in theory it is not clear how electron
correlations induce metallic phase in low dimensional disordered systems.\cite{Kravchenko}  
In the present paper we consider  the interplay of disorder and short range interaction in the MIT. 
Usually, the short range interaction is modelled by the Hubbard interaction.\cite{Hubbard}
Here we take an alternative point of view. Rather than try to study the Hubbard model we take a simpler model,
the spinless Falicov-Kimball model (FKM).\cite{Falicov} 
The relation of the FKM to the Hubbard model is
analogous to the relation between the Ising and Heisenberg models of magnetism. 
The FKM describes itinerant electrons interacting
via a repulsive contact potential with  localized electrons (or ions).
It can also be viewed as a simplified Hubbard model where electrons with down spin are
frozen and do not hop.
Certainly, within the TMT the phase diagram of the Anderson-Mott-Hubbard MIT in the FKM 
and that in the Hubbard model share common features.\cite{Byczuk,Byczuk1}
Moreover, the FKM exhibits a rich phase diagram. In homogeneous phase
it exhibits the Mott-Hubbard type of MIT, although the model does not describe the
Fermi liquid picture. At low temperature different phases with long-range order may exist
depending on the doping and interaction strength.\cite{Kennedy,Uelt,Freericks}  
The FKM can also be incorporated into different models to study various aspects 
of electron correlations, for instance the charge ordered ferromagnetism in manganites.\cite{Tran1,Tran2,Ramak}
In the disordered FKM we can study different realizations of the interplay of disorder and electron
correlations. 
With local disorder the FKM exhibits the Anderson-Mott-Hubbard MIT.\cite{Byczuk1}
It will be studied in the present paper by employing the 
statistical DMFT. After solving of the self-consistent equations of the statistical DMFT we obtain the
distributions of the LDOS. We determine the most probable value of the LDOS and use it to monitor the Anderson
localization.  We find  extended states which occur in the region from weak to intermediate strengths of interaction and disorder. 
For intermediate values of disorder or interaction there is a reentrance of the Anderson
localization. At strong disorder the system is a gapless insulator, while at strong interaction the system is a Mott-Hubbard insulator. We also find at the crossing point from extended states to localization 
the most probable value of the LDOS exhibits a discontinuity.

The plan of the present paper is as follows. In Sec.~II we present the statistical DMFT through
its application to the FKM with local disorder. Numerical results are presented in Sec. III.  
In Sec.~IV conclusion and remarks are presented.

\section{Statistical dynamical mean-field theory}
In this section we describe the statistical DMFT through its application to the FKM with local
disorder. 
The Hamiltonian of the system reads
\begin{eqnarray}
H &=&  \sum_{<i,j>}t_{ij} c^{\dagger}_{i} c^{\null}_{j} + \sum_{i} \varepsilon_i 
c^{\dagger}_{i} c^{\null}_{i} - 
\mu \sum_{i} c^{\dagger}_{i} c^{\null}_{i} \nonumber \\    
&& + E_{f} \sum_{i} f^{\dagger}_{i} f^{\null}_{i} + 
U \sum_{i} c^{\dagger}_{i} c^{\null}_{i} f^{\dagger}_{i} f^{\null}_{i} ,
\label{fkm}
\end{eqnarray} 
where $c^{\dagger}_i$ ($c_i$) and $f^{\dagger}_i$ ($f_i$) is the creation (annihilation) operator of itinerant
 and localized electrons at site $i$, respectively. $t_{ij}$ is the hopping integral of itinerant electrons between
site $i$ and $j$. In the following we take into account only nearest neighbor
hopping, i.e., $t_{ij}=-t$ for nearest neighbor sites, and $t_{ij}=0$ otherwise. We will
use $t$ as the energy unit.
$U$ is the
local interaction of itinerant and localized electrons. $\mu$ is the chemical
potential for itinerant electrons. It controls the electron density. 
$E_f$ is the energy level of localized electrons. It also serves as
the chemical potential of localized electrons and controls the density of localized electrons. 
In the following we will consider only the symmetric
half filling case. It turns out that it is equivalent to $\mu=U/2$ and $E_f=-U/2$.\cite{Kennedy}
$\varepsilon_i$ are independent random variables. They represent  local disorder in the model. 
We will consider the random variables with uniform distribution 
\begin{equation}
P(\varepsilon_i)=\frac{1}{W} \Theta\bigg(\frac{W}{2}-|\varepsilon_i|\bigg) ,
\label{disorder}
\end{equation} 
where $\Theta(x)$ is the step function, and $W$ represents the disorder strength.
When the disorder is absent, model (\ref{fkm}) is the pure FKM. 
In homogeneous phase which occurs at high temperature it exhibits a
Mott-Hubbard MIT.\cite{Kennedy,Uelt,Freericks} 
When the interaction is absent ($U=0$), itinerant and localized electrons are decoupled, and 
the itinerant electron part of model (\ref{fkm}) represents the Anderson
model, and it would exhibit the Anderson localization.\cite{Anderson}
Thus when both disorder and interaction are present, model (\ref{fkm}) would exhibit the complex Anderson-Mott-Hubbard MIT transition at high temperature.\cite{Byczuk1} 
The FKM may be realized
by loading two kinds of fermion atoms with light and heavy masses in optical lattice. The light
atoms play the role of itinerant electrons, while the heavy atoms are kept immobile as 
the localized electrons.  

The main idea of the statistical DMFT is to formulate the DMFT of the system with
a realization of disorder in real space.\cite{Dobrosavljevic} When the disorder is realized, the
local Green function is not homogeneous anymore. In this case we can adopt the 
inhomogeneous DMFT\cite{Tran} for treating the interaction part.
Within the approach the disorder is treated exactly,
while the effects of electron interaction result into the self energy which is
self-consistently calculated by the DMFT equations. However, the self energy
depends on the site index. The electron Green function
can be written in real space  
\begin{equation}
\mathbf{G}(\omega) =[\mathbf{G}_{0}^{-1}(\omega) - \mathbf{\Sigma}(\omega) ]^{-1} ,
\label{dyson} 
\end{equation} 
where $\Sigma_{ij}(\omega)$ is the self energy of the electron Green function
$G_{ij}(\omega)=\langle\langle c_{i} | c^{\dagger}_{j} \rangle\rangle_{\omega} $.
$\mathbf{G}_{0}(\omega)$ is the noninteracting Green function. For the FKM with 
a realization of disorder
$\mathbf{G}_{0}(\omega) =[\omega \delta_{ij} - t_{ij} - \varepsilon_i + \mu \delta_{ij}]^{-1} $.
$\varepsilon_i$ is random variables realized accordingly to the probability
distribution (\ref{disorder}).
Equation (\ref{dyson}) is just the Dyson equation written in the matrix form.
Within the statistical DMFT,  the self energy is approximated by a local function of frequency. 
However,
this local function can vary from site to site, i.e.,
\begin{equation}
\Sigma_{ij}(\omega)=\delta_{ij} \Sigma_{i}(\omega) .
\label{selfidmf}
\end{equation}
The approximation is strictly local.
In infinite dimensions the self energy is purely local. For finite dimensions the approximation
neglects nonlocal correlations. The nonlocal correlations can be systematically incorporated
by cluster extensions of the DMFT.\cite{Maier}
The site dependence of the self energy is generated
via  random variables $\varepsilon_i$. As a consequence the self energy $\Sigma_{i}(\omega)$ is also
 stochastic variables.
With this feature the effective mean field and the local Green function
also are local stochastic variables. The Dyson equation
(\ref{dyson}) shows that 
the self energy gives additional local random contributions to disorder of the system. These contributions
are due to both interaction and the interplay between interaction and disorder. However, in
difference to the random variables $\varepsilon_i$ the contributions are dynamical. They
take into account temporal local quantum fluctuations generated by interaction and disorder. 
They also broaden the random energy levels generated by disorder.
The self energy $\Sigma_{i}(\omega)$ is determined from an effective
single site. Once the effective single site  is solved the
self energy is calculated by the Dyson equation
\begin{equation}
\Sigma_{i}(\omega)=\mathcal{G}_{i}^{-1}(\omega) - G_{i}^{-1}(\omega) ,
\label{dyson1}
\end{equation}
where ${\mathcal{G}}_{i}(\omega)$ is the bare Green function of the effective single site
and represents the effective mean field acting on site $i$. $G_{i}(\omega)$
is the electron Green function of the effective single site. The self-consistent
condition requires that the Green function $G_{i}(\omega)$ of the effective single site
must concise with
the local Green function of the original lattice. i.e.,
\begin{equation}
G_{i}(\omega) = G_{ii}(\omega) .
\label{cons}
\end{equation} 
In the Appendix we show the exact derivation of the self consistent equation for inhomogeneous
systems in infinite dimensions. 
Equations (\ref{dyson})-(\ref{cons}) form the self-consistent system of equations
for the lattice Green function and the self energy. They are principal equations
of the statistical DMFT. 
Since the local Green function is stochastic variables,
the self-consistent equations are naturally stochastic too. In Eq.~(\ref{cons}) the
right hand side is a functional of stochastic variables $G_i(\omega)$, thus the
self-consistent condition (\ref{cons}) generates a stochastic chain of $G_i(\omega)$
via the iteration process. The LDOS is defined as usually
$$
\rho_{i}(\omega)=- \frac{1}{\pi} \; \text{Im} \; G_{ii}(\omega+i\eta) ,
$$
where $\eta=0^{+}$ is an infinitely small positive number.

For the FKM the effective single site problem has the following action
\begin{eqnarray}
S_{i}[c_{i}^{\dagger},c_{i}] &=&
- \int_{0}^{\beta} d\tau d\tau' c_{i}^{\dagger}(\tau) \mathcal{G}_{i}^{-1}(\tau-\tau') c_{i}(\tau')
\nonumber \\
&& + \int_{0}^{\beta} d\tau U (c_{i}^{\dagger}c_{i} f_{i}^{\dagger}f_{i})(\tau)  + \beta E_{f}  f_{i}^{\dagger}f_{i} ,
\end{eqnarray}
where $\beta=1/T$ is the inverse of temperature. The partition function
corresponding to the action is
\begin{equation}
\mathcal{Z}_{i}=\text{Tr}_{f_i} \int \mathcal{D} c_{i}^{\dagger} \mathcal{D} c_{i}
e^{-S_{i}[c_{i}^{\dagger},c_i]} .
\end{equation}
This partition function can be calculated exactly, because the trace over the localized electrons
is independent of the dynamics of itinerant electrons. We obtain\cite{Freericks}
\begin{eqnarray}
\mathcal{Z}_i = 2 \exp\Big[\sum_{n} 
\ln \Big( \frac{\mathcal{G}_{i}^{-1}(i\omega_n)}{i\omega_n} \Big) e^{i\omega_n \eta}\Big]+
\hspace{1.5cm}
\nonumber \\
 2 \exp\Big[-\beta E_{f} + \sum_{n}
\ln \Big( \frac{\mathcal{G}_{i}^{-1}(i\omega_n) - U}{i\omega_n} \Big) e^{i\omega_n \eta}
\Big] ,
\end{eqnarray} 
where $\omega_n=(2n+1)\pi T$ is the Matsubara frequency.
The Green function can directly be calculated from the partition function.
Without difficulty one obtains
\begin{equation}
G_{i}(i\omega_n)=\frac{W_{0i}}{\mathcal{G}_{i}^{-1}(i\omega_n)} +
\frac{W_{1i}}{\mathcal{G}_{i}^{-1}(i\omega_n)-U} ,
\label{glocal}
\end{equation}
where $W_{1i} = f(\widetilde{E}_i)$, $W_{0i}=1-W_{1i}$. Here $f(x)=1/(\exp(\beta x)+1)$ 
is the Fermi-Dirac
distribution function, and
\begin{equation}
\widetilde{E}_{i}= E_{f} + T \sum_{n} \ln
\Big(\frac{1}{1 - U \mathcal{G}_{i}(i\omega_n)} \Big) e^{i\omega_n \eta} .
\label{erenorm}
\end{equation}
Note that the weight factors $W_{0i}$, $W_{1i}$ are not simply a
number. They are functionals of the local Green function. 
One can show that $W_{1i}=\langle f^\dagger_{i} f_i \rangle$ is the density of localized electrons at site $i$.
So far we have obtained the complete solution of the effective single site. This together with
the statistical DMFT equations (\ref{dyson})-(\ref{cons}) fully determine the dynamics of itinerant
electrons with a fixed disorder realization. Within the statistical DMFT, 
the disorder is treated exactly, while the correlation effects are
taken into account through the mean field contributions of the DMFT.    
Once the self-consistent equations of the statistical DMFT are solved we obtain the LDOS. 
With many different
realizations of disorder a data ensemble of the LDOS is obtained. The size of the 
ensemble depends on the lattice size and the number of disorder realizations. 
From the ensemble of the LDOS we can determine its probability distributions as well as
the most probable value of the LDOS. The  most probable value of the LDOS is used to monitor
the MIT driven by disorder and correlation.

The present approach proposes a statistical local description of the MIT in disordered interacting systems.
The statistical aspect is a proper description since it works directly with the ensemble of LDOS and use
the most probable value of the LDOS to monitor the MIT.\cite{Anderson,Anderson1,Anderson2}
The local treatment in the spirit of DMFT neglects nonlocal correlations. This weakness may be serious
in low dimensional systems. However, for the two dimensional FKM nonlocal correlations give nonsignificant
contributions to the DOS in the homogeneous phase.\cite{Hettler} The DMFT calculations for the two-dimensional
FKM also show reasonable results.\cite{Freericks1} Nevertheless, the statistical DMFT can be considered as 
a simplest approach incorporating both disorder and correlation to the MIT.

\section{Numerical results}

In this section we present the numerical results of the statistical DMFT equations.
In general, the matrix inversion in Eq.~(\ref{dyson}) can be performed only
for a finite size lattice. 
We consider a two dimensional square lattice with the linear size $L$.
Thus, we perform numerical calculations for the lattice of size $L$ and finite
number $N_{d}$ of disorder realizations. In particular numerical calculations were performed for
$L=12$ and $N_d=100$. 
We take temperature $T=1$ which is high enough for homogeneous phase and avoiding any 
long-range order. Certainly, the homogeneous phase is insensitive to temperature,
however at low temperature it is unstable against the long-range ordered phase.\cite{Freericks}
The Mott-Hubbard like MIT in the FKM occurs only for the homogeneous phase. 
Strictly speaking, the Anderson localization is well defined only at zero temperature.
Here we study the interplay between the Anderson localization and the Mott-Hubbard MIT
by investigating the single particle Green function
at finite temperature. However, in the absence of long range orders the single particle Green
function is insensitive to temperature. Hence, we can consider the single particle Green function at finite
temperature as it would be at zero temperature. 
The symmetric half filling case with fixed $\mu=U/2$ and 
localized electron density $n_f=1/2$ is considered.
The energy level $E_f$ of localized electrons is determined accordingly by
condition $n_f=\sum_{i} \langle f^\dagger_{i} f_i \rangle /L^2$ for each disorder
realization. We solve the statistical DMFT equations in real frequency by iterations.
The small positive number $\eta=0.01$ is used.
Without disorder the FKM exhibits the Mott-Hubbard MIT with critical value $U_c \approx 4$.
Thus when system is clean the system state is metallic for $U<U_c$, and is insulator
for $U>U_c$.

\subsection{Total DOS and band edge}

\begin{figure}[t]
\vspace{-1.cm}
\begin{center}
\includegraphics[width=0.47\textwidth]{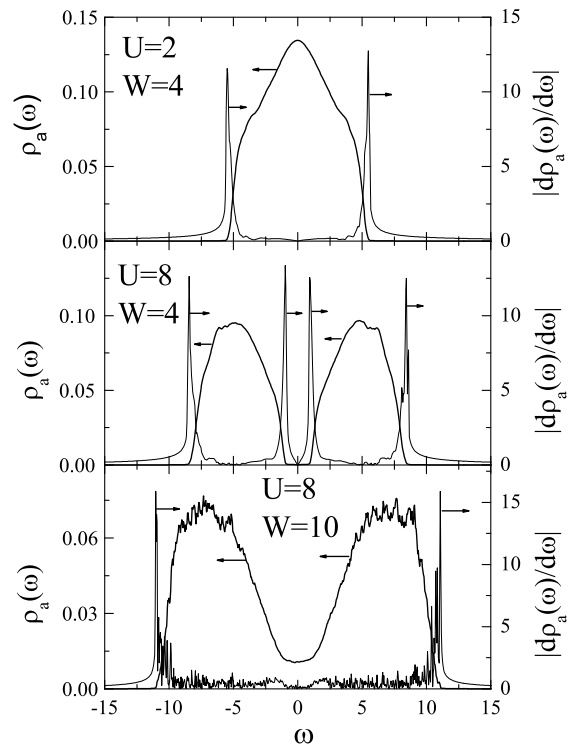}
\end{center}
\vspace{-1cm}
\caption{\label{fig1} The total DOS $\rho_a(\omega)$
and its derivative $| d \rho_a(\omega) / d\omega |$ for
various interactions and disorders.}
\end{figure} 

First we determine the band edge of the system. 
Usually, the band edge is determined from vanishing condition
of the total DOS. 
The total DOS is defined as the arithmetic average of the LDOS 
\begin{equation}
\rho_{a}(\omega)= \frac{1}{N_d L^2} \sum_{\text{disorder}} \sum_{i} \rho_{i}(\omega) .
\end{equation}
However, strictly speaking, in the numerical calculations the total DOS
never vanishes since $\eta=0.01$ was used. 
In Fig.~\ref{fig1} we plot the total DOS and its derivative 
$| d \rho_a(\omega) / d\omega |$.
One can see that at the band edge the total DOS sharply changes and its derivative exhibits a pronounced
peak. We use the position of the peak  to determine
the band edge. The value of the total DOS at the band edge, $\rho_{be}$,  also serves as the cutoff of the DOS.
Below this value $\rho_{be}$ the DOS approximately vanishes. 
For strong interaction and weak disorder, the total DOS opens a gap at the Fermi energy. It is similar to the
Mott-Hubbard insulator in the clean case. For such cases we also use the peaks 
of $| d \rho_a(\omega) / d\omega |$ to determine the band gap.
For strong interaction and strong disorder, the gap opened at the Fermi energy closes. As we will see later,
the system is still localized, but gapless. It is a crossover from the Mott-Hubbard  to
the Anderson insulator by disorder.

\subsection{Anderson-Mott-Hubbard MIT}

\begin{figure}[t]
\vspace{-0.7cm}
\begin{center}
\includegraphics[width=0.5\textwidth]{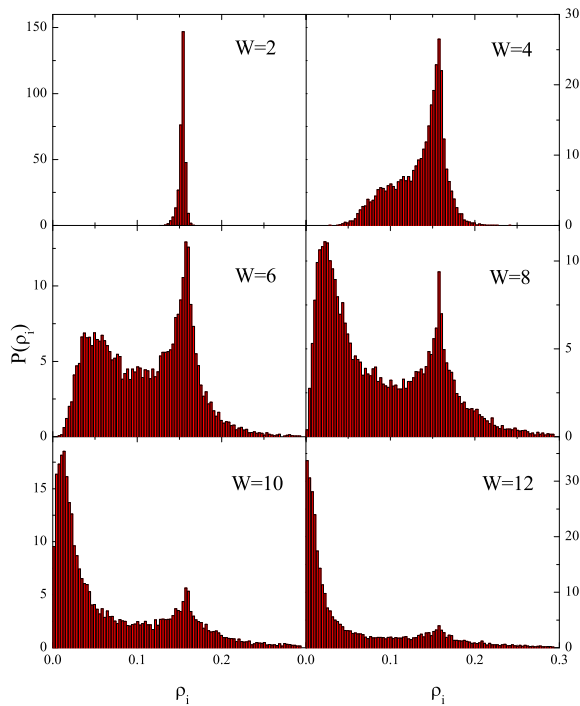}
\end{center}
\vspace{-1.cm}
\caption{\label{fig2} (Color online) Probability distribution $P(\rho_i)$ of the LDOS 
$\rho_i$ at the Fermi energy  
for various disorder $W$ in the weak interaction case ($U=2$).}
\end{figure}

The probability distribution of the LDOS is constructed from statistical data of the LDOS which
is obtained after solving the statistical DMFT equations.
In Fig.~\ref{fig2} we present the histogram of the probability distribution of the LDOS at the Fermi
energy for $U=2$ and various disorders. This value of interaction ($U=2$) corresponds to the
metallic phase in the clean limit.
For weak disorders the probability distribution
has a monomodal structure with a shaped peak at its most probable value.
In this regime the most probable value of the LDOS has a finite value, thus the system 
is in an extended state. This is an unambiguous evidence of the existence of extended states
in the two-dimensional disordered interacting system.
As the disorder strength increases the peak broadens, and then a second peak is developed.
Thus the probability distribution has a bimodal structure. The high value mode is nearly 
fixed almost independently on the disorder strength. The low value mode moves towards to zero value.
Actually, as the disorder strength increases the low value mode becomes the most probable value
of the LDOS, and it approximately vanishes at some value of disorder. The vanishing of the
most probable value of the LDOS is detected in the sense that its value is below 
the density cutoff $\rho_{be}$.
The vanishing of the most probable
value of the LDOS manifests the Anderson localization. Thus the system exhibits a MIT from
extended state to the Anderson localization as the disorder strength increases. 
Perhaps, the bimodal structure of the probability distribution of the LDOS is due to special features
of the FKM. The low value mode is due to the Anderson localization when disorder increases.
The high value mode reflects the non-Fermi liquid
behavior of the FKM in the weak interaction regime. For weak interactions the chemical potential
remains pinned at the effective level of $f$ localized electrons.\cite{Si} In the presence of disorder
most of the LDOS still persist with the pinning property, thus most of the LDOS at the chemical potential
keep the same value. The bimodal structure may be absent in the systems where the Fermi liquid properties are 
maintained.   
 
\begin{figure}[t]
\vspace{-1cm}
\begin{center}
\includegraphics[width=0.5\textwidth]{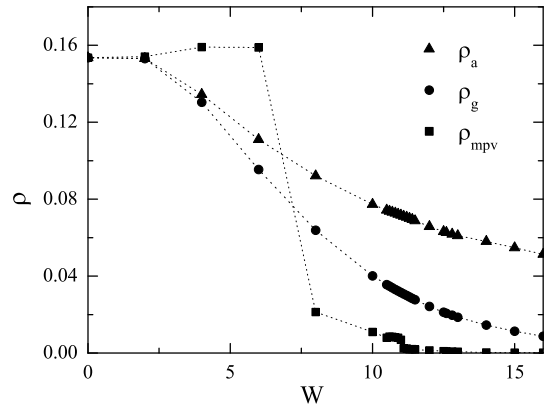}
\end{center}
\vspace{-1cm}
\caption{\label{fig3}  The most probable value $\rho_{\text{mpv}}$, arithmetic
average $\rho_{\text{a}}$ and geometric average $\rho_{\text{g}}$  of the LDOS 
at the Fermi energy via disorder strength $W$ in the weak interaction case 
($U=2$).}
\end{figure} 

In principle, we can determine the most probable value as the value at which 
the probability distribution reaches its global maximum. 
However, the probability distribution is constructed by
histogram, its most probable value is sensitive to the width of histogram bars. 
To avoid the inaccuracy, we determine the most probable value by the half sample mode algorithm.\cite{Bickel}
This algorithm is a fast routine for locating the most probable value of a finite statistical sample. 
The half sample mode algorithm is based on finding the smallest interval that contains half number
of the sample points. The most probable value must lie in the obtained half sample.
Repeat this half sample procedure until obtain the half sample with two or three sample points.
Then one can easily locate the most probable value of the sample.   
In Fig.~\ref{fig3} we plot the most probable value as well as the arithmetic and
geometric average of the LDOS at the Fermi energy for comparison. It shows as
the disorder strength increases the most probable value shifts from the higher value mode to lower
value one. After a critical value of disorder strength ($W_c \approx 11.1$) 
the most probable value approximately vanishes, thus the system changes to the Anderson localized phase.
At the crossing point the most probable value exhibits a discontinuity. The most probable value
of the LDOS is not an order parameter of MIT, since it does not associate with any symmetry breaking, but
the discontinuity of the most probable value of the LDOS may be considered as a sign of the first order
phase transition. 
Figure \ref{fig3} also shows that both the arithmetic and geometric
average of the LDOS never coincide with the most probable value of the LDOS. 
The arithmetic and geometric averages of the LDOS monotonously decrease as the disorder strength increases.
However, the decreases do not necessarily imply the approach to the Anderson localization.\cite{Song}

\begin{figure}[t]
\vspace{-0.7cm}
\begin{center}
\includegraphics[width=0.5\textwidth]{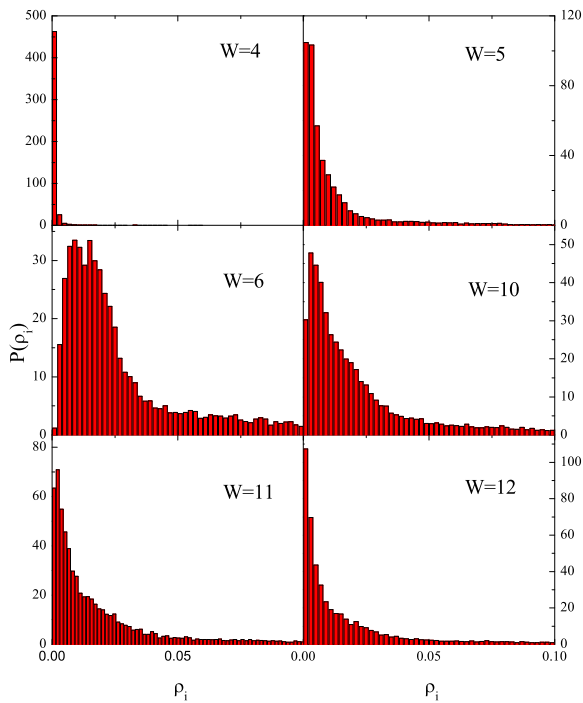}
\end{center}
\vspace{-1cm}
\caption{\label{fig4} (Color online) Probability distribution $P(\rho_i)$ of the LDOS 
$\rho_i$ at the Fermi energy  
for various disorder $W$ in the intermediate interaction case ($U=6$).}
\end{figure}

In Fig.~\ref{fig4} we present the histogram for the probability distribution of the LDOS at the Fermi
energy for various disorders and $U=6$. This value of interaction ($U=6$) corresponds to the
insulator phase in the clean limit. In difference to the weak interaction case, at weak disorders
the probability distribution of the LDOS at the Fermi energy 
has almost a delta function like structure.
It means that almost all LDOS at the Fermi energy vanish. The total DOS opens a gap at the
Fermi energy. It is analogous to the Mott-Hubbard insulator in the clean limit.
As the disorder strength increases, the delta peak broadens, and the probability distribution
of the LDOS shows a long tail. However, the most probable value of the LDOS still approximately
vanishes, while the arithmetic and geometric average are finite. The total DOS now closes the
opened gap at the Fermi energy. The system state is still localized, however gapless. It is
analogous to the Anderson localized phase in the noninteracting limit. We still refer it to
the Anderson localization, although the physics nature may be different. In general,
at finite interaction and finite disorder there are no precise definitions 
of the Mott-Hubbard and Anderson insulating phases.\cite{Byczuk} 
These phases rigorously exist only in the clean or noninteracting limits.
In disordered interacting systems, both phases
are characterized by vanishing of the most probable value of the LDOS at the Fermi energy.
However, the total DOS of the Mott-Hubbard insulator opens a gap at the Fermi energy, while
the one of the Anderson localization is gapless. The scenario of closing the opened gap by disorder
in the strong interaction case can be understood
from the atomic limit.\cite{Aguiar} In the clean atomic limit
each site has two energy levels,
one single occupancy and one double occupancy, separated
by interaction strength $U$. At the half filling, each site is occupied either by one itinerant electron
or by one localized electron. The double occupancy levels remain empty, thus the charge gap is equal to
$U$. When disorder is added, each of these two energy levels is shifted by randomly fluctuating site
energy  $-W/2 < \varepsilon_i < W/2$. For $W<U$ the situation remains unchanged, thus there is 
still a charge gap
for electron excitations. 
When $W>U$, 
the double occupancy levels at some sites may be shifted lower than the single occupancy levels. Thus 
the double occupancy levels of a fraction of sites are either occupied or empty. As a result the charge gap
is closed. The phase may be interpreted as a mixture of Anderson and Mott-Hubbard insulators.   

\begin{figure}[t]
\vspace{-1cm}
\begin{center}
\includegraphics[width=0.5\textwidth]{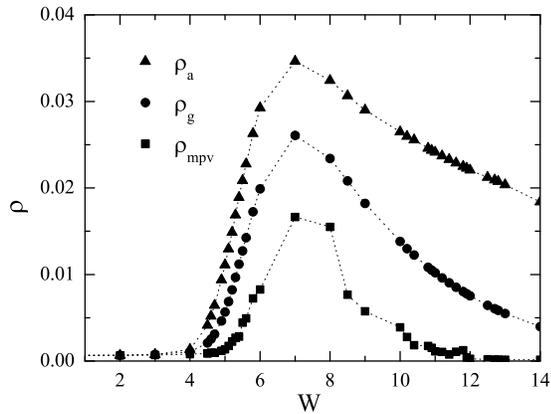}
\end{center}
\vspace{-1cm}
\caption{\label{fig5}  The most probable value $\rho_{\text{mpv}}$, arithmetic
average $\rho_{\text{a}}$ and geometric average $\rho_{\text{g}}$  of the LDOS 
at the Fermi energy via disorder strength $W$ in the strong interaction case  
($U=6$).}
\end{figure} 
 
In contrast to the weak interaction case,
where the probability distribution of the LDOS has the bimodal structure, in the 
intermediate and strong interaction cases
the probability distribution of the LDOS keeps its monomodal structure. 
As the disorder strength increases further, the most probable value
of the LDOS first increases from zero value and then decreases back to zero value.
In the first stage the system changes from the localized state to extended state,
while in the second stage the system changes back from the extended state to localized
state. It is a reentry effect of the Anderson localization.
In Fig.~\ref{fig5} we plot the most probable value,
the arithmetic and geometric average of the LDOS at the Fermi energy as a function of the disorder
strength for $U=6$. In difference to the weak interaction case, 
as the disorder strength increases, the most probable value and
the averages of the LDOS increase from zero value, reach their maximum, and then decrease.  
It also shows that the geometric
average of the LDOS approximately vanishes only in the Mott-Hubbard insulator phase. However in this phase
the arithmetic average and the most probable value of the LDOS approximately vanish too. 
In the Anderson localized
phase only the most probable value of the LDOS vanishes. In the region of intermediate values of 
the disorder strength, the most probable value of the LDOS is finite. This  evidence unambiguously
indicates the existence of extended states for intermediate disorders. It also shows that 
disorder can drive the system from the insulating state to extended one. However, the insulating
state should be the gapless localized state. Disorder cannot drive a Mott-Hubbard
insulator directly to extended state. One may speculate the MIT scenario of the intermediate 
interaction case as a screening of disorder
which leads to close the gap at the Fermi energy, and then the standard scenario of the 
Anderson localization as in the weak interaction case. At the transition point the most probable
value also shows a discontinuity like in the weak interaction case. We can use the discontinuity
to detect the crossing point from extended to localized states.

\begin{figure}[t]
\vspace{-1cm}
\begin{center}
\includegraphics[width=0.5\textwidth]{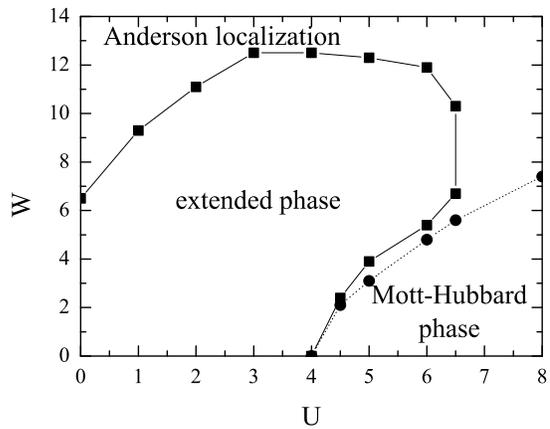}
\end{center}
\vspace{-1cm}
\caption{\label{fig6} Phase diagram for the Mott-Hubbard-Anderson MIT. The dotted line separates
the phase region with finite gap in the total DOS.}
\end{figure} 

In Fig.~\ref{fig6} we plot the phase diagram. It clearly distinguishes three
phase regions. Extended phase is those states that the most probable value of the
LDOS at the Fermi energy is finite. The insulator phase is characterized by the vanishing
of the the most probable value of the LDOS at the Fermi energy. This phase is separated into 
the Anderson localization where the total DOS is gapless and the Mott-Hubbard insulator
where the total DOS opens a gap at the Fermi energy. The Anderson localization
occurs for strong disorder, while the Mott-Hubbard insulator occurs for strong correlation. 
The extended phase appears only for weak and intermediate disorder and correlation. For a weak
disorder, as the interaction increases, the system changes from the extended phase to 
the Anderson localization, and finally to the Mott-Hubbard insulator. For an intermediate
value of disorder, as the interaction increases, the system changes from the Anderson localization
to the extended phase, and then back again to the Anderson localization. This is a reentry
effect of the Anderson localization. 
For a weak interaction as the disorder strength increases the system changes from extended to
localized phase. For intermediate and strong interactions there is a crossover from the
Mott-Hubbard insulator to the Anderson insulator by closing the gap at the Fermi energy by
disorder. For a fixed intermediate interaction there is also the reentry effect of the Anderson
localization when the disorder strength is varied. The phase diagram is qualitatively analogous
to the one calculated within the TMT.\cite{Byczuk1} Although the geometric mean alone does not indicate
the Anderson localization, its fully embedding in the self consistent cycle of the DMFT may describe
the Anderson localization.\cite{Dobrosavljevic1,Byczuk,Byczuk1}
However, within the statistical DMFT there is a discontinuity of the most probable
value of the LDOS at the phase boundary, whereas within the TMT the geometric average is 
continuous at the phase boundary.
The two limiting cases $W=0$ and $U=0$ are special.  
In the clean limit $W=0$, there is the Mott-Hubbard MIT, although the metallic phase is not
a Fermi liquid. The noninteracting limit $U=0$ is controversial in two dimension. 
Our result agrees well
with the real space renormalization group calculations.\cite{Lee1} However, scaling theory
did not find any true metallic behavior in the system.\cite{Abrahams} There is only a crossover
from exponentially to logarithmically localized states. Certainly, the present phase diagram
is constructed from statistics of the LDOS, and the actual transport properties still remain unclear.
Nevertheless, it was demonstrated that the two dimensional
disordered Hubbard model can have a delocalizing effect.\cite{Denteneer,Herbut,Srinivasan,Heidarian}         

\subsection{Finite size effects}

\begin{figure}[t]
\vspace{-0.7cm}
\begin{center}
\includegraphics[width=0.5\textwidth]{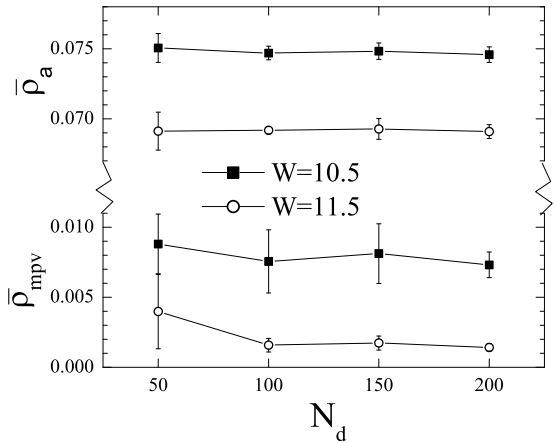}
\end{center}
\vspace{-1cm}
\caption{\label{fig7} The statistical average of the most probable value (mpv) and 
the arithmetic average (a) of the LDOS at the Fermi energy via the number of
disorder realizations. The error bars are their
standard deviations. ($U=2$, $L=12$, $N_b=5$). }
\end{figure} 

\begin{table}[b]

\begin{tabular}{|c||c|c|c|c|c|}
\hline
$L$ & $8$ & $10$ & $12$ & $14$ & $16$ \\ 
\hline
$N_d$ & $240$ & $150$ & $110$ & $80$ & $60$ \\
\hline
$N$ & $15 680$ & $15 000$ & $15 840$ & $15 680$ & $15 360$ \\
\hline
\end{tabular}
\caption{The lattice size $L$, the number of disorder realizations $N_d$,
and the total number $N$, which are used for study of the finite size effects.}
\end{table}

The results of the previous subsections basically permit finite size effects. 
There are two sources of the
finite size effects. One is the finite size $L$ of the lattice, and the second is the finite number $N_d$
of disorder realizations. First, we study the finite size effects of $N_d$. We fix the lattice size $L=12$,
and consider different numbers of disorder realizations. For each $N_d$ we generate $N_b$ different bins of
size $L^2 N_d$ for
disorder realizations. For each bin we calculate the most probable value as well as the arithmetic
average of the LDOS at the Fermi energy. Then we calculate their statistical average
$\overline{\rho}_{\alpha}$ and standard
deviation $\sigma_{\alpha}$, i.e.,
\begin{eqnarray*}
\overline{\rho}_{\alpha} &=& \frac{1}{N_b} \sum_{n=1}^{N_b} \rho_{\alpha}^{(n)} , \\
\sigma_{\alpha}^2 &=& \frac{1}{N_b-1} \sum_{n=1}^{N_b} \big( \rho_{\alpha}^{(n)} - 
\overline{\rho}_{\alpha} \big)^2, 
\end{eqnarray*}  
where $\alpha$ denotes the most probable value (mpv) or the arithmetic average (a).
$\rho_{\alpha}^{(n)}$ is the most probable value or the arithmetic average of the LDOS, obtained
from the nth bin calculations. In Fig.~\ref{fig7} we plot the statistical average
and the standard deviation of the most probable value or the arithmetic average of the LDOS
at the Fermi energy for various $N_d$ and $N_b=5$. We choose two values of disorder nearby
the transition point. One is in the extended phase, and the other is in Anderson localized phase. 
Figure~\ref{fig7} shows the arithmetic average of the LDOS (the total DOS) has little fluctuations
even for $N_d=50$. It almost independent on $N_d$ already from $N_d=50$. The most probable value
of the LDOS fluctuates in the extended phase more strongly than in the localized phase.
Certainly, in the localized phase the most probable value of the LDOS approximately vanishes, that
the fluctuations of the vanishing value is negligible when $N_d$ varies.
The most probable value of the LDOS seems to be reasonable from $N_d=100$. Thus at $N_d=100$
the finite size effects of $N_d$ are small and do not significantly change the results. 

\begin{figure}[t]
\vspace{-0.7cm}
\begin{center}
\includegraphics[width=0.5\textwidth]{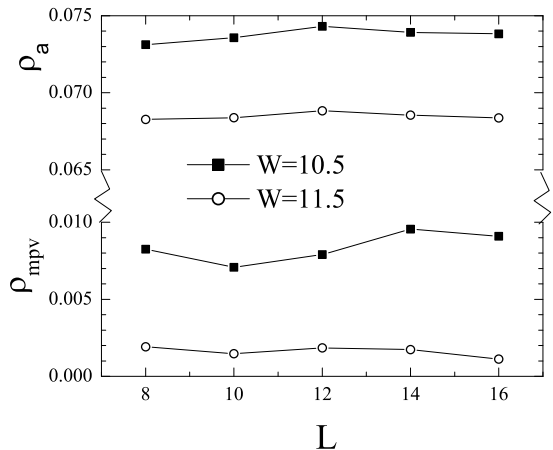}
\end{center}
\vspace{-1cm}
\caption{\label{fig8} The most probable value $\rho_{mpv}$ and 
the arithmetic average $\rho_a$ of the LDOS at the Fermi energy via the lattice
size $L$. The number of disorder realizations for each lattice size is given in
Table~I. ($U=2$). }
\end{figure}

Next, we study the finite size effects of the lattice size. The DMFT calculations for clean systems
show the finite site effects are small and controllable.\cite{Tran} 
In disordered systems, one can notice that the finite
size effects of the ensemble of the LDOS depend mostly on the size of the ensemble, i.e.,
on the number $N=L^2 N_d$. $N$ is the total number of LDOS which are obtained in numerical calculations.
Therefore to study the finite size effects of $L$ alone, when the lattice size $L$ is varied, we have to
change $N_d$ accordingly, that the total number $N$ keeps more or less the same value. In Table~I we present
several values of $L$, and corresponding values of $N_d$ that the total number $N$ is around $15 000$.
We use the parameter values in Table~I for study of the finite size effects of $L$. 
For all cases the whole bin of disorder realizations is 
kept more or less the same. We also choose two values of disorder nearby the transition point. One is in the
extended phase, and the other is in the localized phase. The most probable value and the arithmetic
average of the LDOS at the Fermi energy via the lattice size are plotted in Fig.~\ref{fig8}. It shows
that the arithmetic average of the LDOS is almost independent on the lattice size, at least from
$L=8$. The most probable value of the LDOS also slightly fluctuates as the lattice size varies. These
fluctuations mainly are due to the statistical fluctuations of finite size of disorder realizations.  
As we already showed in Fig.~\ref{fig7}, the statistical fluctuations of the most probable value of the LDOS
in the extended phase is larger than in the localized phase. This feature is consistent with Fig.~\ref{fig8},
where the most probable value of the LDOS fluctuates in the extended phase stronger than in the localized
phase. Both results show that the finite size effects in our study are small and do not change significantly
the picture of MIT. It is interesting to note that the DMFT can be performed for finite size lattices,
and the obtained results are not significantly different from the ones of the thermodynamical limit.\cite{Tran}

\section{Conclusions}

In this paper we have studied the Mott-Hubbard-Anderson MIT in the two-dimensional
FKM with local disorder by the statistical DMFT. Within the statistical DMFT the
correlation effects are resulted into additional dynamical local random variables,
which are self-consistently determined from the local single site dynamics. The
probability distribution and the most probable value of the LDOS are calculated.
The localized phase is detected by vanishing condition of the most probable value
of the LDOS at the Fermi energy. The scenario of the MIT in the system depends on the
interaction. For weak interactions, which correspond to the metallic
phase in the clean limit, the system changes from extended to localized states as the
disorder strength increases. For intermediate interactions, which correspond
to the insulating phase in the clean limit, as the disorder strength increases
the system crosses from the Mott-Hubbard to the Anderson insulator, 
and then it transits to extended states and goes back again to the Anderson localized phase. 
Thus there is a reentrance of the Anderson localization. 
At the crossing point from extended to localized states the most probable value
of the LDOS exhibits a discontinuity. For strong interactions
only localized states exist. There is only a crossover from the Mott-Hubbard to
the Anderson insulator by closing the opened gap at the Fermi energy by disorder. 
The results also confirm
the delocalizing effect in the two dimensional disordered interacting system.
However, the phase diagram was determined only by statistics of the LDOS, and
the actual transport properties of the system still remain unclear. We leave the problem
for further study.

\begin{acknowledgments}

The author would like to thank the Asia Pacific Center
for Theoretical Physics for the hospitality. He also acknowledges
useful discussions with Hanyon Choi and Jaejun Yu.
The author is grateful to thank the Max Planck Institute
for the Physics of Complex Systems at Dresden for sharing computer facilities
where the numerical calculations were performed.
This work was supported by
the Asia Pacific Center for Theoretical Physics, and by the Vietnam National Program
on Basic Research.

\end{acknowledgments}

\appendix
\section{Self consistent equation of the inhomogeneous DMFT in infinite dimensions}

In this Appendix we present the derivation of the self consistent equation of the inhomogeneous DMFT in  infinite dimensions. Formally, we can derive the self consistent equation for inhomogeneous systems in the
same way
as for homogeneous systems.\cite{GKKR} Certainly, the derivation for homogeneous systems is based on the cavity method and the Hilbert transform.\cite{GKKR} 
Since the cavity method for homogeneous systems is also formulated in real space, 
we can follow it closely. First a disorder realization is fixed, and then 
all fermions are traced out except for a single site $l$. In the infinite dimensions we obtain the Green function which represents the effective mean field\cite{GKKR}
\begin{eqnarray}
\mathcal{G}_{l}^{-1}(i\omega_n) &=& i\omega_n + \mu - \varepsilon_l - \Delta_{l}(i\omega_n), 
\label{appd0} \\ 
\Delta_{l}(i\omega_n) &=& \sum_{ij} t_{li} G^{(l)}_{ij}(i\omega_n) t_{jl} ,
\label{appd1}
\end{eqnarray}
where $G_{ij}^{(l)}(i\omega_n)$ is the Green function of the model with site $l$ removed. $\Delta_{l}(i\omega_n)$ can be considered as a hybridization function. 
The cavity Green function can be expressed through the original lattice Green function\cite{GKKR}
\begin{eqnarray}
 G^{(l)}_{ij}(i\omega_n) = G_{ij}(i\omega_n) - 
 \frac{\displaystyle G_{il}(i\omega_n) G_{lj}(i\omega_n)}{\displaystyle G_{ll}(i\omega_n) } .
\label{appd2}
\end{eqnarray}  
Inserting Eq.~(\ref{appd2}) into Eq.~(\ref{appd1}) one obtains
\begin{eqnarray}
\Delta_{l}(i\omega_n) &=&  
  \big[ \mathbf{t} \cdot \mathbf{G}(i\omega_n) \cdot  \mathbf{t} \big]_{ll} 
 \nonumber \\  
 && -
 \frac{\displaystyle \big[ \mathbf{t} \cdot  \mathbf{G}(i\omega_n) \big]_{ll}
 \big[ \mathbf{G}(i\omega_n) \cdot \mathbf{t} \big]_{ll}}
 {\displaystyle G_{ll}(i\omega_n) } , 
\label{appd3}
\end{eqnarray}  
where $\mathbf{t}$ is the hopping matrix. The lattice Green function can be rewritten as
\begin{eqnarray}
\mathbf{G}(i\omega_n) = \big[ \mbox{\boldmath{$\xi$}}(i\omega_n) - \mathbf{t} - \mathbf{\Sigma}(i\omega_n)      
\big]^{-1} ,
\end{eqnarray}
where $\xi_{ij}(i\omega_n) = (i\omega_n+\mu-\varepsilon_i)\delta_{ij}$. In infinite dimensions
the self energy is purely local,\cite{GKKR} hence the self energy matrix $\mathbf{\Sigma}(i\omega_n)$ 
is diagonal. For homogeneous systems
the terms in the right hand side of Eq.~(\ref{appd3}) are calculated by using the Fourier and Hilbert transforms.\cite{GKKR} For inhomogeneous systems the Fourier and Hilbert transforms are replaced by the matrix multiplication and inversion in real space. One can notice that
\begin{eqnarray}
\mathbf{t} \cdot \mathbf{G} \cdot  \mathbf{t} &=&
\mathbf{t} \cdot \mathbf{G} \cdot \big[ \mbox{\boldmath{$\xi$}}  - \mathbf{\Sigma} - \mathbf{G}^{-1}     
\big] \nonumber \\
&=& \mathbf{t} \cdot \mathbf{G} \cdot \big( \mbox{\boldmath{$\xi$}} - \mathbf{\Sigma} \big) - \mathbf{t},
\label{appd4} \\
\mathbf{t} \cdot \mathbf{G} &=& 
 \big[ \mbox{\boldmath{$\xi$}} - \mathbf{\Sigma} - \mathbf{G}^{-1}\big] \cdot \mathbf{G}
\nonumber \\
&=& \big( \mbox{\boldmath{$\xi$}} - \mathbf{\Sigma} \big) \cdot \mathbf{G} -
\mathbf{1} .
\label{appd5}
\end{eqnarray}   
Using relations (\ref{appd4})-(\ref{appd5}), from Eq.~(\ref{appd3}) we obtain
\begin{eqnarray}
\lefteqn{
 \Delta_{l}(i\omega_n) =  -
\big[ \mbox{\boldmath{$\xi$}}(i\omega_n) - \mathbf{\Sigma}(i\omega_n) \big]_{ll} } \nonumber \\
 &+& \Big[\big( \mbox{\boldmath{$\xi$}}(i\omega_n) - \mathbf{\Sigma}(i\omega_n) \big)  \cdot \mathbf{G}(i\omega_n) \cdot
\big( \mbox{\boldmath{$\xi$}}(i\omega_n) - \mathbf{\Sigma}(i\omega_n) \big) \Big]_{ll}\nonumber \\
&-&
\big[ \big( \mbox{\boldmath{$\xi$}}(i\omega_n) - \mathbf{\Sigma}(i\omega_n) \big) \cdot  
\mathbf{G}(i\omega_n) - \mathbf{1} \big]_{ll} \nonumber \\
 && \big[ \mathbf{G}(i\omega_n) \cdot \big( \mbox{\boldmath{$\xi$}}(i\omega_n) - \mathbf{\Sigma}(i\omega_n) \big) - \mathbf{1} \big]_{ll}/
 G_{ll}(i\omega_n) .
 \label{appd6}  
\end{eqnarray}
Here we have used $t_{ll}=0$.
Since both matrices $\mbox{\boldmath{$\xi$}}(i\omega_n)$ and $\mathbf{\Sigma}(i\omega_n)$ are diagonal, 
from Eqs.~(\ref{appd0}) and (\ref{appd6}) we obtain
\begin{eqnarray}
&& \mathcal{G}_{l}^{-1}(i\omega_n) = \Sigma_{l}(i\omega_n) + 1/
 G_{ll}(i\omega_n) .
 \label{appd7}  
\end{eqnarray}
Equation~(\ref{appd7}) is just the self consistent equations (\ref{dyson1})-(\ref{cons}). 
It is exact in infinite dimensions for any inhomogeneous system. The above derivation of the
self consistent equation can be considered as a generalization of the homogeneous one. In infinite
dimensions, the sum in Eq.~(\ref{appd1}) runs over infinitely many neighbouring sites, that the
hybridization function is replaced by its average value. As a consequence, all spatial fluctuations
of the environment surrounding a given site are suppressed, thus the Anderson localization is 
prohibited.\cite{Dobrosavljevic2}
At infinite dimensions the Anderson localization is absent
in any disordered interacting system.

\end{document}